\documentstyle[aps, 12pt]{revtex}
\begin{document}
\draft
\title{Quantum Key Distribution and Quantum Authentication Based on Entangled State}
\author{Bao-Sen Shi, Jian Li, Jin-Ming Liu, Xiao-Feng Fan and Guang-Can Guo}
\address{Laboratory of Quantum Communication and Quantum Computation, Department of\\
Physics\\
University of Science and Technology of China\\
Hefei, 230026, P. R. China}
\maketitle

\begin{abstract}
Using the previously shared Einstein-Podolsky-Rosen pairs, a proposal which
can be used to distribute a quantum key and identify the user's
identification simultaneously is presented. In this scheme, two local
unitary operations and the Bell state measurement are used. Combined with
quantum memories, a cryptographic network is proposed. One advantage is no
classical communication is needed, which make the scheme more secure. The
secure analysis of this scheme is shown.\newline
\end{abstract}


\section{Introducation}

Perfectly secure communication between two parties can be achieved if they
share beforehand a common random sequence of bits (a key), so how to
distribute a secret key is very important for secure communication. In
classical cryptography, there is nothing to prevent an eavesdropper from
monitoring the key distribution channel passively without being caught by
the legitimate users. In quantum cryptography, quantum key distribution
(QKD) [1] has been proposed as a new solution to this problem. QKD is a
technique that permits two parties, who share no secret information
initially, to establish a shared secret sequence of bits, its security is
based on quantum law, such as ``no-cloning'' theorem. Since the publication
of the BB84 protocol [1], QKD has developed into a well-understood
application of quantum mechanics to cryptography. Many theoretical schemes
have been proposed [2-5] and many experiments have been done [6-10].

Typically, QKD schemes depend either on an unjammable classical
communication channel or on authentication of the classical communication by
classical methods. Generally, the assumption that the classical
communication channel is unjammable seems unpractical, so the key
authentication is very important for the security. Recently, several quantum
authentication (QA) schemes [11-16] have been proposed. $Dusek$ $et$ $al$
[11] presented a proposal based on the combination of classical
identification procedure and QKD. In the proposals [12,13], the parties
initially share entanglement. Another kind of proposal is based on
entanglement catalyst [14,15]. In these schemes, either unjammable classical
communication is needed [11-15], or legal users previously share a sequence
of secret bits [11, 13]. In this paper, we present a scheme, by which, QKD
and QA can be realized simultaneously. In this scheme, some
Einstein-Podolsky-Rosen (EPR) pairs are previously shared by two parties,
called Alice and Bob. When they want to establish a sequence of secret keys,
any party, for example Bob, sends his particles back to Alice after he does
one of two local unitary operations $I$ and $\sigma _x$ randomly, where $%
I=\left[ 
\begin{array}{cc}
1 & 0 \\ 
0 & 1
\end{array}
\right] ,$ $\sigma _x=\left[ 
\begin{array}{cc}
0 & 1 \\ 
1 & 0
\end{array}
\right] .$ After having received particles from Bob, Alice makes a Bell
state measurement on two particles belong in original EPR pair. By this way,
Alice and Bob can share a sequence of keys, at the same time, Alice and Bob
can identify the identification of each other. In this scheme, only
previously shared EPR pairs are needed. During the QKD and QA, compared to
other schemes, no classical communication and previously shared secret bits
are needed if the quantum channel is error-free. Furthermore, this scheme
can be used to transmit information directly because of no discarded bits
during the transmission in the case of the error-free channel.

If this scheme is combined with the quantum memories [17], QKD between any
pair of parties can be realized. No classical communication is needed in the
network in the case of error-free channel.

In section II, we present a new two-party quantum cryptographic scheme,
which can be realize simultaneously QKD and QA. In section III, we present a
quantum network based on this scheme with the addition of quantum memories.
In section IV, we give a brief discussion and conclusion.

\section{A two-party QKD and QA scheme}

Suppose that Alice and Bob have previously shared K pairs entangled states in

\begin{equation}
\Psi ^{-}=\frac 1{\sqrt{2}}[\left| 01\right\rangle -\left| 10\right\rangle ],
\end{equation}
where the first particle is held by Alice and the second particle is held by
Bob. QKD and QA in this scheme consist of the following steps.

1. Bob performs randomly one of two local unitary operations $I$ and $\sigma
_x$ on his particle in each EPR pair.

2. Bob sends his particle back to Alice.

3. After having received this particle, Alice does a Bell state measurement
on the particle from Bob and the particle from herself, these two particle
initially belong in the same EPR pair $\Psi ^{-}$ .

If Bob performs the unitary operation $I$ on the particle belong in him, the
state $\Psi ^{-}$ holds unchanged. If the unitary operation performed by Bob
is $\sigma _x$, the state $\Psi ^{-}$ will be transformed into state $\Phi
^{-}$, where $\Phi ^{-}=\frac 1{\sqrt{2}}[\left| 00\right\rangle -\left|
11\right\rangle ],$ which is another Bell state. The other two Bell states
are $\Psi ^{+}=\frac 1{\sqrt{2}}[\left| 01\right\rangle +\left|
10\right\rangle ]$ and $\Phi ^{+}=\frac 1{\sqrt{2}}[\left| 00\right\rangle
+\left| 11\right\rangle ]$ respectively. So when Alice does a Bell state
measurement, she should only get the result of state $\Phi ^{-}$ or $\Psi
^{-}$ if no eavesdropper exists.

4. After having completed transmission, Alice and Bob let state $\Psi ^{-}$
and the unitary operation $I$ correspond to binary ``0'', the state $\Phi
^{-}$ and the unitary operation $\sigma _x$ correspond to binary ``1'', then
they can share a key, at the same time, identification of Bob is
authenticated. In order to identify Alice, this scheme needs a small
revision. We let Bob send a certain number of particles back to Alice,
alternatively, let Alice send next certain number of particles back to Bob.
Of course, Alice also does one of two unitary operations $I$ or $\sigma _x$
on the particle belong in her before she sends the particle back to Bob. Bob
does the same measurement to identify Alice. The number is decided before
transmission by Alice and Bob, and it is public to everyone, not secret. By
this way, we can realize the QKD and QA simultaneously.

Obviously, no classical communication and no previously shared bits are
needed in this scheme, which may make our scheme more secure. Furthermore,
no discarded bits makes this scheme transmit the information directly in the
case of error-free channel.

Now, we discuss the security of our scheme. Firstly, we analyze the
intercept/resend strategy. When Alice or Bob sends particle back to each
other, an eavesdropper Eve may intercept this particle and resend a fake
particle instead according to her measurement result. For example, when Bob
sends his particle back to Alice, Eve intercepts it . Because the state of
particle belong in Bob is

\begin{equation}
\rho _B=Tr_A\rho _{\Psi ^{-}}=Tr_A\rho _{\Phi ^{-}}=\frac 12\{\left|
0\right\rangle \left\langle 0\right| +\left| 1\right\rangle \left\langle
1\right| \},
\end{equation}
Eve can not get any information by this way. If Eve resends a fake particle
to Alice, for example, this fake particle is in the state $\phi _E=c\left|
0\right\rangle +d\left| 1\right\rangle ,$ where, $\left| c\right| ^2+\left|
d\right| ^2=1.$When Alice receives this particle, she does a Bell state
measurement on this fake particle and the particle belong in herself. The
state of the fake particle and the particle of Alice's is

\begin{equation}
\rho _{AE}=\frac 12\{\left| 0\right\rangle \left\langle 0\right| +\left|
1\right\rangle \left\langle 1\right| \}\otimes \{c^2\left| 0\right\rangle
\left\langle 0\right| +cd^{*}\left| 0\right\rangle \left\langle 1\right|
+c^{*}d\left| 1\right\rangle \left\langle 0\right| +d^2\left| 1\right\rangle
\left\langle 1\right| \}.
\end{equation}
If Alice does a Bell state measurement on these two particles, she will get
any one of four Bell states with equal probability 1/4. If Alice gets the
result state $\Psi ^{+}$ or $\Phi ^{+}$, she can conclude that eavesdropper
exists.

Now, we discuss another strategy. Suppose when Bob sends his particle back
to Alice, Eve performs a CONTROLLED-NOT (CNOT) operation . The control
particle is the particle from Bob, the target is an ancillary particle $%
\left| 0\right\rangle _E$ owned by Eve. If the control particle is $\left|
0\right\rangle $, the target particle holds unchanged. If the control
particle is $\left| 1\right\rangle $, the target particle flips. After CNOT
operation, the state of three particle is

\begin{equation}
\Psi _{AB}^{-}\otimes \left| 0\right\rangle _E\stackrel{CNOT}{\rightarrow }%
\frac 1{\sqrt{2}}[\left| 011\right\rangle -\left| 100\right\rangle ]_{ABE}
\end{equation}
or

\begin{equation}
\Phi _{AB}^{-}\otimes \left| 0\right\rangle _E\stackrel{CNOT}{\rightarrow }%
\frac 1{\sqrt{2}}[\left| 000\right\rangle -\left| 111\right\rangle ]_{ABE}.
\end{equation}

Then Eve holds the ancillary particle and let the particle from Bob to pass
to Alice. After having received the particle from Bob, Alice does the Bell
state measurement . If Bob makes $I$ unitary on his particle, then Alice can
get the state $\Psi ^{-}$ or the state $\Psi ^{+}$ with the probability 50\%
respectively. As we knows, no result $\Psi ^{+}$ can be got if no
eavesdropper exists. If Bob makes the unitary operation $\sigma _x$ on his
particle, then Alice will get the result $\Phi ^{-}$ or $\Phi ^{+}$ with the
probability 1/2 respectively. The result $\Phi ^{+}$ should not appear if no
eavesdropper exists. So, if Eve uses this strategy, she can not get any
information at all and legal users will detect her.

Can Eve learn any information from the ancillary particle? the answer is no.
Whether the unitary operation is $I$ or $\sigma _x$, the state of the
ancillary particle is the same, which is $\rho _{A=}\frac 12\{\left|
0\right\rangle \left\langle 0\right| +\left| 1\right\rangle \left\langle
1\right| \}$, so Eve can not get any information from the ancillary particle.

According to the above analysis, our scheme is secure if the error-free
channel is used.

\section{A quantum network based on this scheme with the addition of quantum
memories}

In this section, we combine the two-party QKD scheme with the use of quantum
memories to present a quantum cryptographic network. Our scheme is similar
to the scheme of Ref. [17], in which a quantum file owned by the center is
needed. Of course, some differences exist between these two schemes. Our
scheme can be summaried as the follows:

1. In the preparation step the user prepares L pairs EPR states $\Psi ^{-}$
and sends one particle of each EPR pair to the center, holds the other
particle of each EPR pair by himself. The center keeps these particles in a
quantum file without measuring them.

2. When users Alice and Bob wish to obtain a common secure key, they ask the
center to create correlation between two strings, one of Alice and one of
Bob. The center perform the Bell state measurement on each pair of qubit,
which realize the entanglement swapping. After that, Alice and Bob share a
EPR pair. What the Bell state will be obtained depends on the result of the
Bell state measurement performed by the center. For example, if the Bell
state measurement result performed by the center is $\Psi ^{-}$ , the Bell
state shared by Alice and Bob is $\Psi ^{-}$ too. Then Alice and Bob can
realize the QKD with the previous scheme in section II. The table I
summaries the results obtained by the center, state shared by Alice and Bob,
the unitary operations and Bell state measurement performed by Alice or Bob,
corresponding binary number.

\[
Table\text{ }I 
\]

\begin{tabular}{|c|c|c|c|c|}
\hline
center state & Alice and Bob state & unitary operation & Bell state
measurement & binary number \\ \hline
$\Psi ^{-}$ & $\Psi ^{-}$ & $I$ or $\sigma _x$ & $\Psi ^{-}$ or $\Phi ^{-}$
& 0 or 1 \\ \hline
$\Psi ^{+}$ & $\Psi ^{+}$ & $I$ or $\sigma _x$ & $\Psi ^{+}$ or $\Phi ^{+}$
& 0 or 1 \\ \hline
$\Phi ^{-}$ & $\Phi ^{-}$ & $I$ or $i\sigma _y$ & $\Phi ^{-}$ or $\Psi ^{+}$
& 0 or 1 \\ \hline
$\Phi ^{+}$ & $\Phi ^{+}$ & $I$ or $i\sigma _y$ & $\Phi ^{+}$ or $\Psi ^{-}$
& 0 or 1 \\ \hline
\end{tabular}

The reason that when state shared by Alice and Bob is $\Phi ^{+}$ or $\Phi
^{-}$ , the unitary operation is $i\sigma _y$ instead of $\sigma _x$ will be
shown in the following.

3. An honest center, which perform the correct entanglement swapping does
not get any information on the strings.

4. If a cheating center ( or any eavesdropper who might have had access to
the quantum files), who modifies the allowed sates, unavoidably introduces
error between the two strings. The analysis about this case is similar to
Ref. [17]. There is another case, in which the center does not tell the
correct Bell state measurement result, for example, he tells Alice and Bob
that the result is $\Phi ^{+}$ instead of correct result $\Psi ^{+}$. In
this case, if the unitary operation $I$ or $\sigma _x$ is used to distribute
QKD, the strategy of the center will be successful, because $\Psi ^{+}%
\stackrel{I}{\rightarrow }\Psi ^{+}$, $\Psi ^{+}\stackrel{\sigma _x}{%
\rightarrow }\Phi ^{+},$ $\Phi ^{+}\stackrel{I}{\rightarrow }\Phi ^{+},\Phi
^{+}\stackrel{\sigma _x}{\rightarrow }\Psi ^{+}.$ Obviously, if the center
uses this strategy, then Alice and Bob can not detect him and will share a
reversed binary number. In order to resolve this problem, we use the unitary
operator $i\sigma _y$ instead of $\sigma _x$. If the center is so careless
that he make a Bell measurement on the pair consisted of particle from Alice
and particle from other user, not from Bob, by this scheme, this mistake can
be detected by Alice and Bob. Because Alice and Bob do not share the EPR
pair, when Alice or Bob does a Bell state measurement on their particle, the
result which should not appear will be obtained with certain probability.
There is another complicated strategy, in which, the center let unlegal user
named Charley as a legal user Bob, and make Alice and Charley share EPR
pairs instead of Alice and Bob. Besides, the center let Charley as legal
user Alice and make Charley and Bob share another EPR pairs. When Alice
(Bob) sends back her (his) particles to Bob (Alice), Charley can intercept
them and make a Bell state measurement, then resends his particles back to
Bob (Alice) according to his result of the measurement. Obviously, our
scheme is useless to this kind of strategy. This means the authentication in
the network depends mainly on the center.

\section{Discussions and Conclusions}

A two-party scheme can distribute a key and identify the user's
identification simultaneously.. One advantage of this scheme is only
previously shared EPR pairs are needed, no classical communication and
previously shared classical secret key are needed. Besides, this scheme can
be used to transmit the information directly in the case of error-free
channel. Based on this scheme with the addition of the quantum memories, a
network similar to the scheme of Ref. [17] can be realized. Contrast to the
Ref. [17], no classical communication is needed. Besides, no qubit is
discarded in the case of error-free channel. One of another advantage is no
information is included in EPR pair before QKD. One disadvantage is quantum
channel is needed, and any user needs to keep quantum state, which make this
scheme complicated. Our scheme seems like a public-key cryptsystem. The
unitary operations can be regraded as public key and the secret key is EPR
correlation. One problem is how to distribute EPR pairs, one possible way is
present in Ref. [18]. In practice, one disadvantage of our scheme is how to
realize a completely Bell state measurement. Very recently, {\it Kim et a}l
[19] proposed a scheme which can solve this problem. Another disadvantage is
the storage of the EPR pair , which is also serious problem in quantum
computer. Now, it is already possible to keep the quantum state in the spin
of the ions for more than 10 Min. [20]and in principle it is possible to
keep them for years. So, Our scheme may be implementable in practice.

This paper is supported by the National Natural Science Foundation of the
China (Grant. No 69907005).


\begin{references}
\bibitem{}  C.H.Bennett, G.Brassard, in proc IEEE int conf on Computers
System,and Signal Processing, Bangalore (IEEE, New York, 1984)P175

\bibitem{}  A.K.Ekert, Phys.Rev.Lett 67 (1991)661

\bibitem{}  C.H.Bennett, Phys.Rev.Lett 68 (1992)3121

\bibitem{}  G-C Guo and B-S Shi, Phys. Lett. A, 256,109 (1999)

\bibitem{}  B.-S Shi, Y.-K Jiang and G.-C Guo, Appl. Phys. B, 70, 415, (2000)

\bibitem{}  W. T. Buttler, et. al., Phys. Rev. Lett., 81, 3283, (1998)

\bibitem{}  A. Muller, et. al., Appl. Phys. Lett., 70, 793, (1997)

\bibitem{}  P. D. Townsend, Elec. Lett., Vol. 33, No. 3, 188, (1997)

\bibitem{}  A. V. Sergienke, et. al, Phys. Rev, A., 60, R2622, (1999)

\bibitem{}  J. D. Franson and B. C. Jacobs, Elec. Lett., 31, 232, (1995)

\bibitem{}  M. Dusek, et. al., Phys. Rev. A., 60, 149, (1999)

\bibitem{}  G. Zeng and W. Zhang, Phys. Rev. A, 61, 022303, (2000)

\bibitem{}  Y.-S Zhang, C.-F. Li and G.-C Guo, e-print arxiv quant-ph 0008044

\bibitem{}  H. Barnum, e-print arxiv quant-ph 9910072

\bibitem{}  J. G. Jensa and R. Schack, e-print arxiv quant-ph 0003104

\bibitem{}  D. Liunggren, M. Bourennane and A. Karlsson, Phys. Rev. A, 62,
022305, (2000)

\bibitem{}  E. Biham, B, Huttner and T. Mor, Phys. Rev. A., 54, (1996) 2651

\bibitem{}  Chui-Ping Yang and Guang-Can Guo, Phys. Rev. A., 59, (1999) 4217

\bibitem{}  Y-H Kim, et. al., e-print arxiv quant-ph 0010046

\bibitem{}  J. J. Bollinger, et. al., IEEE. Trans. Inst. Meas. , 40, 126,
(1997)
\end{references}
\end{document}